\documentclass{aa} 
\usepackage[varg]{txfonts}
\usepackage{natbib}
\usepackage{graphicx}

\begin{document}

\title{Early gray dust formation in the type IIn SN 2005ip}

\author{Ann-Sofie Bak Nielsen\inst{1,2}
  \and Jens Hjorth\inst{2}
  \and Christa Gall\inst{2} 
    } 

\institute{Leiden Observatory, Leiden University, Niels Bohrweg 2, 
2333 CA Leiden, The Netherlands
  \and 
Dark Cosmology Centre, Niels Bohr Institute, University of Copenhagen, 
Juliane Maries Vej 30, DK-2100 Copenhagen, \O, Denmark
}  

\date{Received 14 October 2016 / Accepted XXX XXX 2017}

\abstract{The physical characteristics of dust formed in supernovae is poorly 
known. In this paper, we investigate the extinction properties of dust formed 
in the type IIn SN 2005ip. The observed light curves of SN 2005ip all exhibit 
a sudden drop around 50 days after discovery. This has been attributed to dust 
formation in the dense circumstellar medium. We modeled the intrinsic light 
curves in six optical bands, adopting a theoretical model for the luminosity 
evolution of supernovae interacting with their circumstellar material. From 
the difference between the observed and intrinsic light curves, we calculated extinction 
curves as a function of time. The total-to-selective extinction 
ratio, $R_V$, was determined from the extinction in the $B$ and $V$ bands. 
The resulting extinction, $A_V$, increases monotonically up to about 1 mag, 
150 days after discovery. The inferred $R_V$ value also increases slightly 
with time, but appears constant in the range 4.5--8, beyond 100 days after 
discovery. The analysis confirms that dust is likely formed in SN 2005ip, 
starting about two months after explosion. The high value of $R_V$, that is, 
gray dust, suggests dust properties different from  of the Milky Way. 
While this result hinges on the assumed theoretical intrinsic light curve 
evolution, it is encouraging that the fitted light curves are as expected for 
standard ejecta and circumstellar medium density structures.}

\maketitle

\section{Introduction}
How do the properties of dust formed in supernovae (SNe) compare to those 
observed in the interstellar medium (ISM) of galaxies? The total-to-selective 
extinction ratio, $R_V$, 
is an empirical indicator of the dependency of the extinction 
upon wavelength
along the line of sight to a reddened object; higher values of 
$R_V$ result in a more gray (flat) extinction curve. The ratio 
$R_V$ is known to depend on the sizes, shapes, and composition
of the dust grains.
The average value of 
$R_V$ is around 3 in Local Group galaxies; the Milky Way (MW), and the Large 
and Small Magellanic Clouds have average $R_{V}$ values of 3.1, 3.41, and 2.7, 
respectively \citep{Gordon_2003ApJ...594..279G}. In dense Galactic molecular 
clouds, higher values of $R_{V}\sim$ 4--6 \citep{1989ccm,ISM} are observed. 
Recently, $R_V = 4.5 \pm 0.2$ was measured in 30 Doradus, indicating a gray 
component in the extinction law, attributed to a larger percentage of large 
grains \citep{Marchi_2016MNRAS.455.4373D}.

Core-collapse SNe originate from massive stars with masses more than 
$8-10M_\odot$ \citep{Heger_2003ApJ...591..288H,
Colgate_1966ApJ...143..626C,Ibeling_2013ApJ...765L..43I}, and are divided 
into various classes, based on their spectral and photometric properties 
(e.g., \citealt{Filippenko_1997ARA&A..35..309F}). Type IIn SNe are 
characterized by emission lines composed of a narrow velocity component on 
top of an intermediate or broad velocity width component. The narrow emission 
originates from the slow moving circumstellar medium (CSM) 
\citep{Schlegel_1990MNRAS.244..269S}, which is material shed by the progenitor 
via mass loss in the later stages of the life of a star 
\citep[e.g.,][]{2010Fox,2011Gall,Smith_2016arXiv161202006S}. 
The peak brightnesses of type IIn SNe span a wide range 
\cite[4--5 mag;][]{2012Stritzinger}. The observed light curves are determined 
by the morphology of the CSM and by the progenitor star \citep{2010vanMarle}. 
For example, the shape of the light curves of type IIn SNe depend on the 
initial radius of the star, the ejecta mass and the explosion energy of the 
SNe, and the density structures of the ejecta and CSM
\citep{2010vanMarle,Moriya_2013MNRAS.435.1520M,2015A&A...580A.131T}. 
The range of potential progenitors is large and may include any star with a 
significant pre-SN eruption, for example, red supergiants or luminous blue variables 
(LBVs) \citep{2011Gall,2012Stritzinger,VanDyk_2013AJ....145..118V}. 

\citet{1991supe.conf...82L} noted a shallow optical extinction curve inferred 
from the light curves of SN 1987A. More recently, \cite{2014Gall} examined the 
formation of dust in the luminous type IIn SN 2010jl 
\citep{Newton_2010CBET.2532....1N} and found $R_V\approx6.4$. While this was 
inferred from significant attenuation of the red wings and corresponding 
blueshifts of the centroids of the hydrogen emission lines, no broadband 
color changes were detected in the light curves.

SN 2005ip was discovered on 2005 November 5.2 UT in the Scd galaxy NGC 2906 
\citep{2005Boles_Nakano,2012Stritzinger}. The measured redshift of the host 
galaxy is 0.00714, corresponding to a luminosity distance of about 30 Mpc 
\citep{Vaucouleurs_1991rc3..book.....D,2009SmithB}. SN 2005ip exhibits clear 
evidence of narrow line emission arising from the CSM and is thus classified 
as a type IIn SN \citep{2009SmithB}. The exact date of the SN explosion is 
unknown, but is suggested to be 8--10 days prior to discovery 
\citep{2009SmithB}. The progenitor is believed to be a star with dense CSM. 
There are two suggestions for a progenitor star of SN 2005ip. One is an 
extreme red supergiant, as suggested by \citet{2009SmithB}, because the 
inferred clumpy medium in the CSM is similar to small scale clumps found 
in the red hypergiant star VY Canis Majoris. The other suggestion is an LBV 
star owing to a high mass-loss rate, which is similar to what has been observed 
in other LBVs \citep{2010Fox,Katsuda_2014ApJ...780..184K}. 

There is evidence pointing to dust forming in a cool dense shell (CDS) behind 
the forward shock passing through the dense CSM as early as 50--100 days past
explosion in some SNe, such as SN 1998S, SN 2006jd, or SN 2010jl 
\citep{Pozzo_2004MNRAS.352..457P,2012Stritzinger,2014Gall}. Indeed, SN 2005ip 
shows signs of such early dust formation, as inferred from the attenuation of 
the red wing of the H$\alpha$ line and the observed drop in the light 
curves of the optical bands \citep{2012Stritzinger}.

Here we determine the extinction properties, notably $R_V$, for the dust 
formed early in SN 2005ip. We examine its well-sampled light curves 
\citep{2012Stritzinger} and model the intrinsic light curves in section 
\ref{powerlaw}. From the theoretical and observed light curves we can 
determine the extinction curve that allows the determination of $R_V$. In 
section \ref{discussion} we discuss the possible origins of a shallow 
extinction curve.

\section{Light curves and extinction measurement}\label{powerlaw}

\subsection{Data}\label{sec:data}

\begin{figure}[h!]
\centering
\vspace{-2.2cm}
\includegraphics[width=1.0\linewidth]{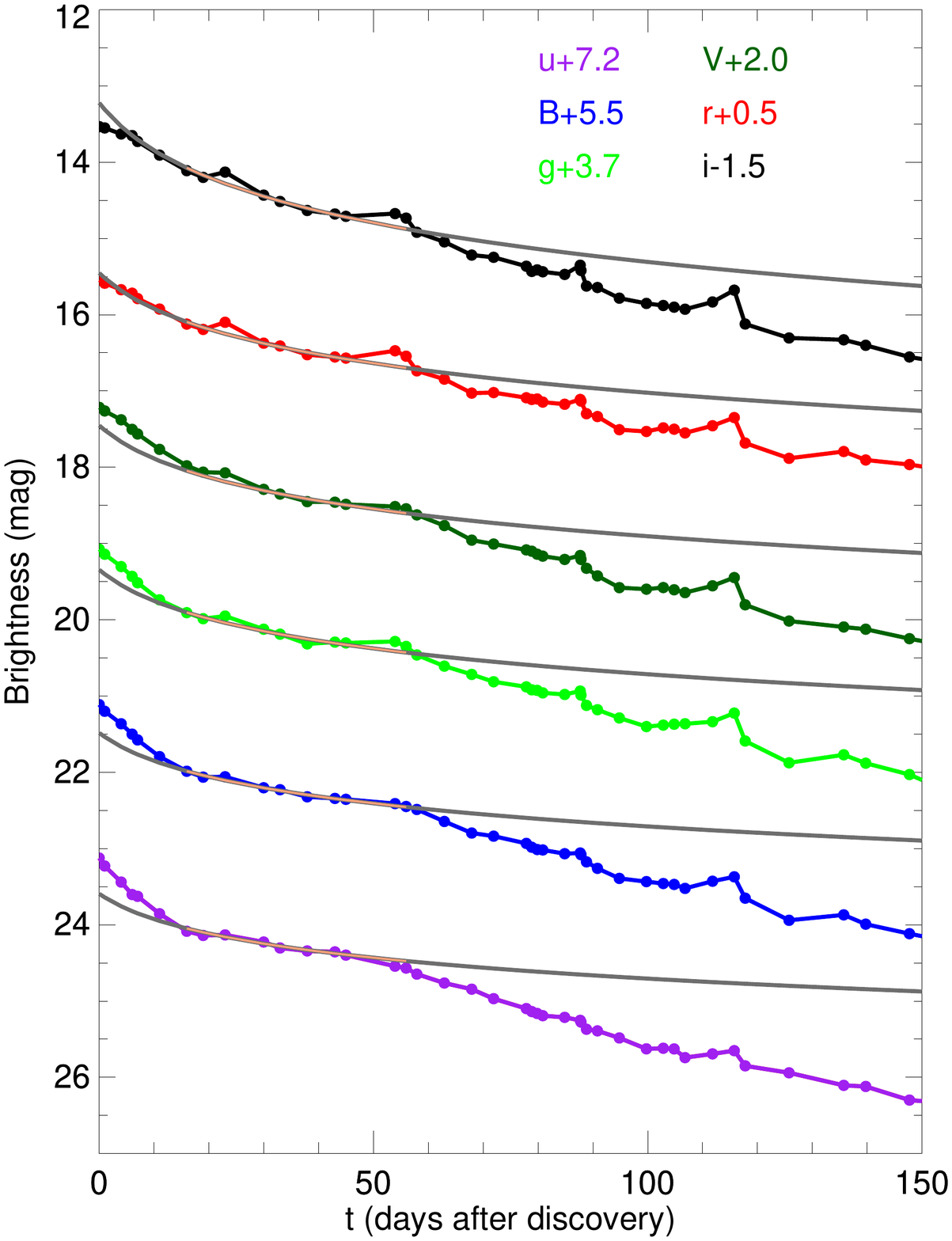}
\caption{Optical light curves for the type IIn SN 2005ip. The connected 
colored filled circles represent the observational data 
\citep{2012Stritzinger}. The light curves exhibit a marked decrease after 
about +50 days, which is an indication of early dust formation. The gray 
curves represent fits of a theoretical model (Equation~\ref{eq:magnitude}) to 
the light curves between $+$23 and $+$47 days. The fitting range is 
highlighted. The model intrinsic light curves are all clearly brighter than 
the observed light curves from $+$50 days and onward.} 
\label{fig:LC_all_log_new}
\end{figure}

The data we use here consist of optical broadband photometry of SN 2005ip from 
the Carnegie Supernova Project 
\citep{2012Stritzinger,Hamuy_2006PASP..118....2H}. 
Photometry in the near-infrared bands is not useful for our purposes because 
it is affected by hot dust emission. As is evident from 
Fig.~\ref{fig:LC_all_log_new}, a clear drop in the light curves is observed in 
all six bands at around 50 days past discovery, which is ascribed to dust 
formation in the SN, for e, by \citet{2012Stritzinger}. Throughout this paper 
the time after the day of discovery is written as $t = +x$ days.

\subsection{Intrinsic light curves}\label{sec:intrinsic_LC}

We fit a theoretical model for the intrinsic light curve to the observed light 
curves of SN 2005ip between $+$23 and $+$47 days, i.e., before dust formation 
sets in; see Fig.~\ref{fig:LC_all_log_new}. It is then assumed that the model 
intrinsic light curves are valid beyond +47 days. 
 
The intrinsic light curves are parametrized according to the theoretical model 
for the light curves of CSM interaction-powered SNe of 
\citet{Moriya_2013MNRAS.435.1520M}. Specifically, for an ejecta density 
$\rho_{\rm ejecta} \propto r^{-n}$ \citep[e.g.,][]{Matzner_1999ApJ...510..379M} 
and a CSM density $\rho_{\rm CSM} \propto r^{-s}$, 
\citet{Moriya_2013MNRAS.435.1520M} found that the bolometric luminosity decays
as a power law as a function of time since explosion $t'$,
\begin{equation}\label{eq:magnitude}
m_{\lambda,\rm intrinsic} (t') =
m_{\lambda, \rm intrinsic} ({\rm 1\ day})-2.5 \alpha \log(t'),
\end{equation}
where we adopt $t' = t-9$ days \citep{2009SmithB}, and
\begin{equation}\label{eq:alpha}
\alpha=\frac{6s-15+2n-ns}{n-s}.
\end{equation}
For example, for a red supergiant progenitor with $n\approx 12$ and a steady 
wind CSM density structure ($s=2$), $\alpha = -0.3$.

Following \citet{Moriya_2013MNRAS.435.1520M}, we leave $\alpha$ as a free 
parameter, and fit Equation~(\ref{eq:magnitude}) to the six individual light 
curves. We choose the temporal fitting range to be +23 to +47 days to be in 
the range of validity of the model (i.e., well past peak), and to avoid being 
affected by the effects of dust attenuation. 
\citet{Moriya_2013MNRAS.435.1520M} considered their model valid up to at least 
$+$200 days while \citet{2009SmithB} noted a remarkable flattening of the 
light curve beyond $+$160 days. To be conservative, we here assume that the 
extrapolated model fits are valid up to $+$150 days. One thousand synthetic datasets 
were created in a 3$\sigma$ Gaussian distribution around the original dataset, 
using a bootstrap Monte Carlo method \citep{MC}. The best fits shown in 
Fig.~\ref{fig:LC_all_log_new} are seen to overshoot the observed light curves 
beyond $+$50 days. We verified that our results do not depend sensitively on 
the adopted time of explosion or the fitting range. 

The fits to the individual light curves yield values of $\alpha$ ranging from 
$-0.77$ to $-0.41$. As shown in Fig.~\ref{fig:ext_B1_all} the best-fit values 
of $\alpha$ are wavelength dependent, where $-\alpha$ increases with 
wavelength. This suggests some color evolution of the SN light, where the 
SN becomes slightly bluer with time, before the conspicuous onset of dust 
formation at around $+$50 days, consistent with what was found by, for example, 
\citet{2013A&A...555A..10T}. Within the uncertainties, however, the majority 
of the light curves are consistent with a value of around 
$\alpha = -0.5\pm0.1$. For example, $\alpha = -0.49$ would correspond to, 
for example, ($n,s$) = (12,2.3), which is broadly consistent with a red supergiant progenitor.

\subsection{Extinction curves}\label{extinction}

The extinction as a function of time for each photometric band is defined as 
the difference between the observed light curve and the model intrinsic light 
curve, as
\begin{equation}\label{eq:ext}
A_\lambda (t) = m_{\lambda,\rm observed} (t) -m_{\lambda,\rm intrinsic} (t).
\end{equation}
To minimize the effects of the apparent bumps in the light curves we smoothed 
the extinction versus\ time curves by fitting fifth order polynomials. 

\begin{figure}[h!]
\centering
\vspace{-2.2cm}
\includegraphics[width=1.00\linewidth]{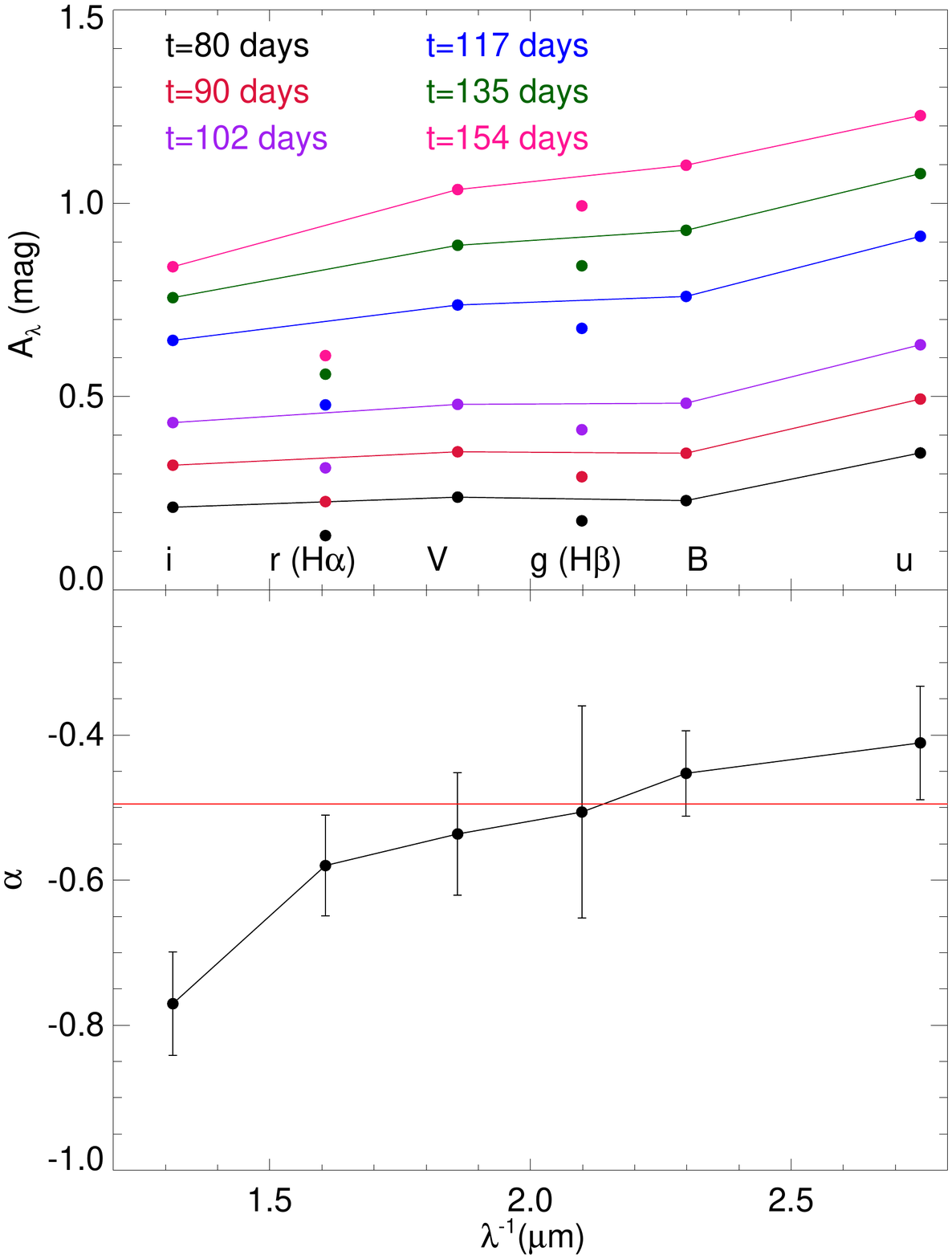}
\caption{Lower panel: Inferred values of $\alpha$ as a function of inverse 
wavelength. The value of $\alpha$ appears to depend slightly on wavelength.
An indicative value of $\alpha=-0.49$ is shown as a red horizontal line. 
Upper panel: Inferred extinction curves at different times. The optical bands 
used in the study are indicated. The $g$ band is affected by H$\beta$ emission and the 
$r$ band is affected by H$\alpha$ emission. The extinction is seen to increase 
with time and the extinction curves are fairly flat, suggesting gray 
extinction.}
\label{fig:ext_B1_all}
\end{figure}

The resulting extinction curves are shown in Fig.~\ref{fig:ext_B1_all} at 
different epochs. The extinction increases with time, maintaining a fairly 
constant shape of the extinction curve as a function of inverse wavelength.  
There are conspicuous dips in the extinction curves in photometric bands 
containing the strong emission lines H$\alpha$ ($r$ band) or H$\beta$ 
($g$ band). 

The total-to-selective extinction is given by the extinction in a specific 
band and the color excess between two photometric bands \citep{ISM},
\begin{equation}\label{eq:RV_E}
R_V=\frac{A_V}{E(B-V)},
\end{equation}
where $E(B-V)$ is the color excess between the $B$ and $V$ bands, $A_V$ is 
the extinction in the $V$ band, and $R_V$ is the total-to-selective extinction 
ratio with respect to the $V$ band. In order not to bias our results, we 
decided not to fit parametric extinction curves to the data (upper panel of 
Fig.~2), but rather calculated $R_V$ directly from Eq.~(\ref{eq:RV_E}), that is,
using the smoothed extinctions as a function of time in the $B$ and $V$ bands.

\begin{figure}[h!]
\centering
\vspace{-2.2cm}
\includegraphics[width=1.00\linewidth]{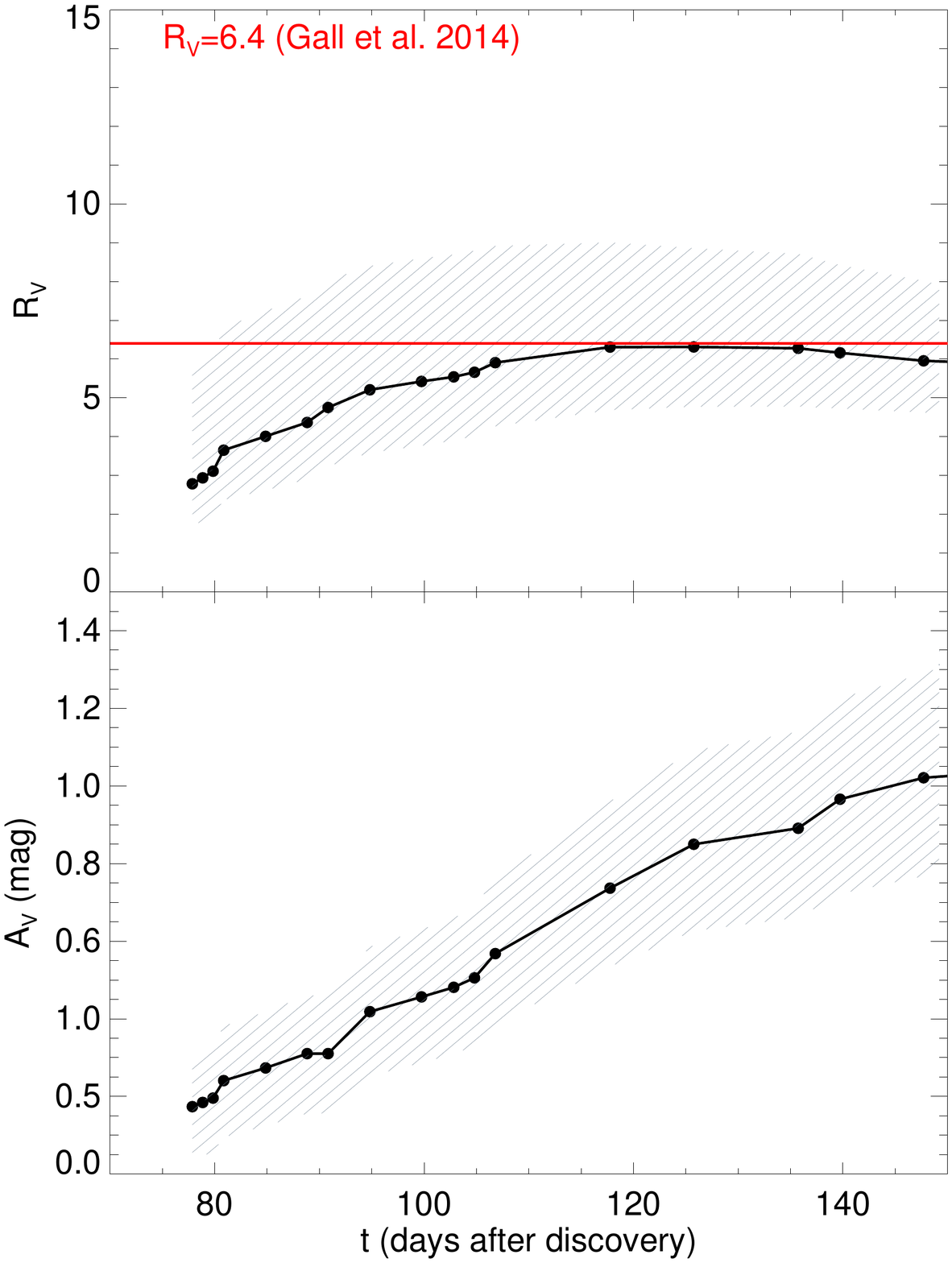}
\caption{Lower panel: $A_V$ (connected filled circles) increases monotonically, 
as expected in the case of dust formation. The shaded area represents the 
1$\sigma$ error region. Upper panel: $R_V$ as a function of time. The value of 
$R_{V}=6.4$ found for SN 2010jl \citep{2014Gall} is indicated as a red 
horizontal line.}
\label{fig:Av_T_all_logpoly}
\end{figure}

Figure~\ref{fig:Av_T_all_logpoly} shows $A_V$ and $R_V$ as a function of time.
The shaded areas represent the 1$\sigma$ error range in the values of $A_V$ and 
$R_V$, as obtained from the Monte Carlo resampling of the data. The curves 
represent the peak values of the distribution. The $A_V$ value increases monotonically 
with time up to a level of about 1 mag at $+$150 days. 

The mean value of $R_V$ appears to increase with time, although within errors 
it is consistent with being in the range 4.5--6.5 after $+$80 days and in the 
range 4.5--8 beyond $+$100 days; $R_V \approx 4.5$ represents the maximum 
lower limit at $+$120 days while $R_V\approx 8$ represents the minimum upper 
limit at $+$150 days. This range is consistent with the value of 
$R_V\approx6.4$ found by \citet{2014Gall} for SN 2010jl.

\section{Discussion}\label{discussion}

The shallow extinction curves obtained here for SN 2005ip are intriguing. 
The derived $R_{V}$ values are higher than the average values for the ISM of 
the MW, SMC, or LMC, but are consistent with MW $R_{V}$ values of some 
individual sight lines \citep{1989ccm}. Furthermore, the derived $R_{V}$ 
values for SN 2005ip are similar to that of SN 2010jl \citep{2014Gall}, which 
is also a type IIn SN with a dense CSM. Typically, a wavelength dependent 
extinction curve reflects the dust grain composition and grain size 
distribution -- properties, which can be obtained by fitting detailed dust 
models to the extinction curve. However, for SN 2005ip, we do not have 
sufficient information to perform such fits. In what follows, we briefly 
discuss the possible origin of such a shallow SN extinction curve, that is, a 
high $R_{V}$.

In SN 2005ip, the early dust formation is believed to occur in the dense CSM
\citep{2009SmithB}. The primary formation site is the CDS, which is formed 
behind the forward shock propagating through the CSM. The CDS environment%
\footnote{We note that the CDS in other types of SNe, such as SNe~IIP 
may be less massive than for SNe IIn and hence less favorable 
for dust to efficiently form and grow. For example, \citet{2011ApJ...732..109M} 
find that in the IIP SN 2004dj only $\sim$10$^{-6}$ M$_\odot$ of dust formed, 
with grain sizes of $\sim$0.2 $\mu$m in a CDS with a mass of
$\sim$10$^{-4}$ M$_\odot$.}
has very different physical and chemical properties than the 
SN ejecta for which most dust formation models are developed 
\citep[e.g.,][]{2011Gall,Sarangi_2015A&A...575A..95S,Mauney_2015ApJ...800...30M,
Lazetti_2016ApJ...817..134L}. Such a formation site has been suggested to 
facilitate the early dust formation in, for example, SN 2006jc  
\citep{2008ApJ...680..568S}, SN 2010jl \citep{2014Gall}, or SN 2011ja 
\citep{Andrews_2016MNRAS.457.3241A}. Very high densities can be reached in 
such a CDS leading to efficient cooling of the material to temperatures 
$\lesssim$ 2000 K, which are suitable for dust to nucleate and possibly grow to 
larger sizes through accretion.

{\em Large dust grains}:
Large $\mu m$-sized grains can lead to shallow extinction curves
\citep{Maiolino_2001A&A...365...37M, Nozawa_2013ApJ...770...27N}. 
Large dust grains have been claimed in the remnant of SN 1987A 
\citep{Wesson_2015MNRAS.446.2089W,2016MNRAS.456.1269B}, although these formed 
later than about three years after the explosion. Larger dust grains, 
depending on the exposure of the dust grains to sputtering and hard radiation 
\citep{Slavin_2004ApJ...614..796S}, are more likely to survive throughout the 
SN remnant phase and thus enrich the ISM 
\citep{2014Gall,Wesson_2015MNRAS.446.2089W,Owen_2015ApJ...801..141O}.

{\em Small, nonstandard dust grains}:
The wavelength dependence of extinction curves is also determined by the 
wavelength dependent emissivity of the relevant dust species. Typically, the 
emissivity of carbonaceous and some silicate dust is highly wavelength 
dependent for small grains in the optical wavelength regime 
\citep{Rouleau_1991ApJ...377..526R,Bladh_2012A&A...546A..76B}, yielding 
nearly constant $R_{V}$ $\approx$ 2.4 for graphite, $R_{V}$ $\approx$ 3.8 for 
carbonaceous dust, and $R_{V}$ $\approx$ 4.5 for silicate dust with grains 
$\lesssim$ 0.01 $\mu$m. For grains of sizes between 0.01 and 0.1 $\mu$m, 
depending on the species, $R_{V}$ can become as low as $\sim$ 1.0.

However, SiO$_2$ exhibits very little to no wavelength dependence in the 
optical wavelength regime for small grains 
\citep[e.g.,][and references therein]{Bladh_2012A&A...546A..76B}. 
This may be an attractive alternative to large carbonaceous grains because 
small grains are easier to nucleate and grow, provided they are shielded from 
UV radiation in an optically thick region \citep{2014arXiv1407.7856K}.
However, the absorption efficiency of SiO$_2$ is about 4 orders of 
magnitude less than that of carbonaceous dust \citep{Bladh_2012A&A...546A..76B}.

{\em Clumpy medium}:
An alternative mechanism for producing a shallow extinction curve may involve 
the semiopaque high-density clumps found in the CSM of SN 2005ip 
\citep{2009SmithB}. If the clumps in the medium were completely opaque, 
the result would be a wavelength independent extinction curve, which we do not 
observe. However, if the clumps were not completely opaque they could lead to 
the shallow extinction curves we see. A similar effect was explored in the 
case of the clumpy ejecta of SN 1987A \citep{2007MNRAS.375..753E}.

\section{Conclusions}\label{conclusion}
SN 2005ip showed evidence of early dust formation in the densely sampled
multicolor optical light curves \citep{2012Stritzinger}. In this paper we 
have used these data to obtain the total-to-selective extinction ratio, $R_V$, 
of dust formed in SN 2005ip. In doing so, we used a theoretical model for the 
light curves of SNe interacting with their circumstellar material 
\citep{Moriya_2013MNRAS.435.1520M} to model the intrinsic light curves and fit the early light curves (i.e., prior to the cutoff around 
$+$50 days). We obtained the extinction as a function of time in each band as 
the difference between the observed light curve and the extrapolated model 
intrinsic light curve up to $+$150 days. From this we derived the 
total-to-selective extinction $R_V$. 

We find that the obtained extinction curves are shallow; the curves are described by values 
of $4.5<R_V<8$ from $+$100 days and onward, which is higher than typically 
found in the ISM of galaxies such as the MW ($R_V\sim3.1$). The high values of 
$R_V$ are consistent with that derived for SN 2010j ($R_V\approx6.4$), where 
dust model fits to the extinction curves suggested large grain formation  
\citep{2014Gall}. 

The advantage of the large grain scenario is the increased survivability of 
the grains throughout the SN remnant phase. Unfortunately, it is challenging 
to form $\mu$m-sized carbonaceous grains on such short timescales. 
It may be easier to form smaller nm-sized grains in high-density
regions, although it is unclear if the required species, whose emissivities 
exhibit little wavelength dependence in the optical regime,
are sufficiently abundant in the CSM. Clumpy material in the CSM may
contribute to flattening the extinction curves.

We also demonstrated that taking the extinction of the dust formed into 
account is essential to derive the intrinsic value of $\alpha$ of the 
\citet{Moriya_2013MNRAS.435.1520M} model. The value we derive for $\alpha$ 
appears to be wavelength dependent, but is consistent with $\alpha \approx -0.49$, 
as would be expected for a standard ejecta density 
$\rho_{\rm ejecta} \propto r^{-12}$ profile 
\citep[][]{Matzner_1999ApJ...510..379M} and a steady wind density structure, 
$\rho_{\rm CSM} \propto r^{-2.3}$. This is in marked contrast to the values 
around $\alpha\approx-1$ found by \citet{Moriya_2013MNRAS.435.1520M}, who included the parts of the light curves that are affected by significant dust attenuation.

The method devised here to determine SN dust properties from multicolor light 
curves may be used to obtain new insight into SN dust formation processes from
future high-quality SN light curves, for example, from the Large Synoptic Survey Telescope.

\begin{acknowledgements}
A.S.B.N. acknowledges funding from a NWO Vidi fellowship. 
J.H. was supported by a VILLUM FONDEN Investigator grant (project number 16599). 
C.G. acknowledges funding from the Carlsberg Foundation.
The authors would like to thank A.G.G.M. Tielens and the anonymous referee for helpful comments.
\end{acknowledgements}

\bibliographystyle{aa} 
\bibliography{bib2} 

\begin{thebibliography}{45}
\expandafter\ifx\csname natexlab\endcsname\relax\def\natexlab#1{#1}\fi

\bibitem[{{Andrews} {et~al.}(2016){Andrews}, {Krafton}, {Clayton}, {Montiel},
  {Wesson}, {Sugerman}, {Barlow}, {Matsuura}, \&
  {Drass}}]{Andrews_2016MNRAS.457.3241A}
{Andrews}, J.~E., {Krafton}, K.~M., {Clayton}, G.~C., {et~al.} 2016, \mnras,
  457, 3241

\bibitem[{{Bevan} \& {Barlow}(2016)}]{2016MNRAS.456.1269B}
{Bevan}, A. \& {Barlow}, M.~J. 2016, \mnras, 456, 1269

\bibitem[{{Bladh} \& {H{\"o}fner}(2012)}]{Bladh_2012A&A...546A..76B}
{Bladh}, S. \& {H{\"o}fner}, S. 2012, \aap, 546, A76

\bibitem[{{Boles} {et~al.}(2005){Boles}, {Nakano}, \&
  {Itagaki}}]{2005Boles_Nakano}
{Boles}, T., {Nakano}, S., \& {Itagaki}, K. 2005, Central Bureau Electronic
  Telegrams, 275, 1

\bibitem[{{Cardelli} {et~al.}(1989){Cardelli}, {Clayton}, \&
  {Mathis}}]{1989ccm}
{Cardelli}, J.~A., {Clayton}, G.~C., \& {Mathis}, J.~S. 1989, \apj, 345, 245

\bibitem[{{Colgate} \& {White}(1966)}]{Colgate_1966ApJ...143..626C}
{Colgate}, S.~A. \& {White}, R.~H. 1966, \apj, 143, 626

\bibitem[{{De Marchi} {et~al.}(2016){De Marchi}, {Panagia}, {Sabbi}, {Lennon},
  {Anderson}, {van der Marel}, {Cignoni}, {Grebel}, {Larsen}, {Zaritsky},
  {Zeidler}, {Gouliermis}, \& {Aloisi}}]{Marchi_2016MNRAS.455.4373D}
{De Marchi}, G., {Panagia}, N., {Sabbi}, E., {et~al.} 2016, \mnras, 455, 4373

\bibitem[{{de Vaucouleurs} {et~al.}(1991){de Vaucouleurs}, {de Vaucouleurs},
  {Corwin}, {Buta}, {Paturel}, \&
  {Fouqu{\'e}}}]{Vaucouleurs_1991rc3..book.....D}
{de Vaucouleurs}, G., {de Vaucouleurs}, A., {Corwin}, Jr., H.~G., {et~al.}
  1991, {Third Reference Catalogue of Bright Galaxies. Volume I: Explanations
  and references. Volume II: Data for galaxies between 0$^{h}$ and 12$^{h}$.
  Volume III: Data for galaxies between 12$^{h}$ and 24$^{h}$.}

\bibitem[{{Ercolano} {et~al.}(2007){Ercolano}, {Barlow}, \&
  {Sugerman}}]{2007MNRAS.375..753E}
{Ercolano}, B., {Barlow}, M.~J., \& {Sugerman}, B.~E.~K. 2007, \mnras, 375, 753

\bibitem[{{Filippenko}(1997)}]{Filippenko_1997ARA&A..35..309F}
{Filippenko}, A.~V. 1997, \araa, 35, 309

\bibitem[{{Fox} {et~al.}(2010){Fox}, {Chevalier}, {Dwek}, {Skrutskie},
  {Sugerman}, \& {Leisenring}}]{2010Fox}
{Fox}, O.~D., {Chevalier}, R.~A., {Dwek}, E., {et~al.} 2010, \apj, 725, 1768

\bibitem[{{Gall} {et~al.}(2011){Gall}, {Hjorth}, \& {Andersen}}]{2011Gall}
{Gall}, C., {Hjorth}, J., \& {Andersen}, A.~C. 2011, \aapr, 19, 43

\bibitem[{{Gall} {et~al.}(2014){Gall}, {Hjorth}, {Watson}, {Dwek}, {Maund},
  {Fox}, {Leloudas}, {Malesani}, \& {Day-Jones}}]{2014Gall}
{Gall}, C., {Hjorth}, J., {Watson}, D., {et~al.} 2014, \nat, 511, 326

\bibitem[{{Gordon} {et~al.}(2003){Gordon}, {Clayton}, {Misselt}, {Landolt}, \&
  {Wolff}}]{Gordon_2003ApJ...594..279G}
{Gordon}, K.~D., {Clayton}, G.~C., {Misselt}, K.~A., {Landolt}, A.~U., \&
  {Wolff}, M.~J. 2003, \apj, 594, 279

\bibitem[{{Hamuy} {et~al.}(2006){Hamuy}, {Folatelli}, {Morrell}, {Phillips},
  {Suntzeff}, {Persson}, {Roth}, {Gonzalez}, {Krzeminski}, {Contreras},
  {Freedman}, {Murphy}, {Madore}, {Wyatt}, {Maza}, {Filippenko}, {Li}, \&
  {Pinto}}]{Hamuy_2006PASP..118....2H}
{Hamuy}, M., {Folatelli}, G., {Morrell}, N.~I., {et~al.} 2006, \pasp, 118, 2

\bibitem[{{Heger} {et~al.}(2003){Heger}, {Fryer}, {Woosley}, {Langer}, \&
  {Hartmann}}]{Heger_2003ApJ...591..288H}
{Heger}, A., {Fryer}, C.~L., {Woosley}, S.~E., {Langer}, N., \& {Hartmann},
  D.~H. 2003, \apj, 591, 288

\bibitem[{{Ibeling} \& {Heger}(2013)}]{Ibeling_2013ApJ...765L..43I}
{Ibeling}, D. \& {Heger}, A. 2013, \apjl, 765, L43

\bibitem[{{Katsuda} {et~al.}(2014){Katsuda}, {Maeda}, {Nozawa}, {Pooley}, \&
  {Immler}}]{Katsuda_2014ApJ...780..184K}
{Katsuda}, S., {Maeda}, K., {Nozawa}, T., {Pooley}, D., \& {Immler}, S. 2014,
  \apj, 780, 184

\bibitem[{{Kochanek}(2014)}]{2014arXiv1407.7856K}
{Kochanek}, C.~S. 2014, arXiv:1407.7856

\bibitem[{{Lazzati} \& {Heger}(2016)}]{Lazetti_2016ApJ...817..134L}
{Lazzati}, D. \& {Heger}, A. 2016, \apj, 817, 134

\bibitem[{{Lucy} {et~al.}(1991){Lucy}, {Danziger}, {Gouiffes}, \&
  {Bouchet}}]{1991supe.conf...82L}
{Lucy}, L.~B., {Danziger}, I.~J., {Gouiffes}, C., \& {Bouchet}, P. 1991, in
  Supernovae, ed. S.~E. {Woosley}, 82

\bibitem[{{Maiolino} {et~al.}(2001){Maiolino}, {Marconi}, \&
  {Oliva}}]{Maiolino_2001A&A...365...37M}
{Maiolino}, R., {Marconi}, A., \& {Oliva}, E. 2001, \aap, 365, 37

\bibitem[{{Matzner} \& {McKee}(1999)}]{Matzner_1999ApJ...510..379M}
{Matzner}, C.~D. \& {McKee}, C.~F. 1999, \apj, 510, 379

\bibitem[{{Mauney} {et~al.}(2015){Mauney}, {Buongiorno Nardelli}, \&
  {Lazzati}}]{Mauney_2015ApJ...800...30M}
{Mauney}, C., {Buongiorno Nardelli}, M., \& {Lazzati}, D. 2015, \apj, 800, 30

\bibitem[{{Meikle} {et~al.}(2011){Meikle}, {Kotak}, {Farrah}, {Mattila}, {Van
  Dyk}, {Andersen}, {Fesen}, {Filippenko}, {Foley}, {Fransson}, {Gerardy},
  {H{\"o}flich}, {Lundqvist}, {Pozzo}, {Sollerman}, \&
  {Wheeler}}]{2011ApJ...732..109M}
{Meikle}, W.~P.~S., {Kotak}, R., {Farrah}, D., {et~al.} 2011, \apj, 732, 109

\bibitem[{{Moriya} {et~al.}(2013){Moriya}, {Maeda}, {Taddia}, {Sollerman},
  {Blinnikov}, \& {Sorokina}}]{Moriya_2013MNRAS.435.1520M}
{Moriya}, T.~J., {Maeda}, K., {Taddia}, F., {et~al.} 2013, \mnras, 435, 1520

\bibitem[{{Newton} \& {Puckett}(2010)}]{Newton_2010CBET.2532....1N}
{Newton}, J. \& {Puckett}, T. 2010, Central Bureau Electronic Telegrams, 2532

\bibitem[{{Nozawa} \& {Fukugita}(2013)}]{Nozawa_2013ApJ...770...27N}
{Nozawa}, T. \& {Fukugita}, M. 2013, \apj, 770, 27

\bibitem[{{Owen} \& {Barlow}(2015)}]{Owen_2015ApJ...801..141O}
{Owen}, P.~J. \& {Barlow}, M.~J. 2015, \apj, 801, 141

\bibitem[{{Pozzo} {et~al.}(2004){Pozzo}, {Meikle}, {Fassia}, {Geballe},
  {Lundqvist}, {Chugai}, \& {Sollerman}}]{Pozzo_2004MNRAS.352..457P}
{Pozzo}, M., {Meikle}, W.~P.~S., {Fassia}, A., {et~al.} 2004, \mnras, 352, 457

\bibitem[{Press {et~al.}(2001)Press, Teukolsky, Vetterling, \& Flanney}]{MC}
Press, W.~H., Teukolsky, S.~A., Vetterling, W.~T., \& Flanney, B.~T. 2001,
  Numercial Recipes in Fortran 77, The art of scientific computing (Cambridge
  University Press)

\bibitem[{{Rouleau} \& {Martin}(1991)}]{Rouleau_1991ApJ...377..526R}
{Rouleau}, F. \& {Martin}, P.~G. 1991, \apj, 377, 526

\bibitem[{{Sarangi} \& {Cherchneff}(2015)}]{Sarangi_2015A&A...575A..95S}
{Sarangi}, A. \& {Cherchneff}, I. 2015, \aap, 575, A95

\bibitem[{{Schlegel}(1990)}]{Schlegel_1990MNRAS.244..269S}
{Schlegel}, E.~M. 1990, \mnras, 244, 269

\bibitem[{{Slavin} {et~al.}(2004){Slavin}, {Jones}, \&
  {Tielens}}]{Slavin_2004ApJ...614..796S}
{Slavin}, J.~D., {Jones}, A.~P., \& {Tielens}, A.~G.~G.~M. 2004, \apj, 614, 796

\bibitem[{{Smith}(2016)}]{Smith_2016arXiv161202006S}
{Smith}, N. 2016, arXiv:1612.02006

\bibitem[{{Smith} {et~al.}(2008){Smith}, {Foley}, \&
  {Filippenko}}]{2008ApJ...680..568S}
{Smith}, N., {Foley}, R.~J., \& {Filippenko}, A.~V. 2008, \apj, 680, 568

\bibitem[{{Smith} {et~al.}(2009){Smith}, {Silverman}, {Chornock}, {Filippenko},
  {Wang}, {Li}, {Ganeshalingam}, {Foley}, {Rex}, \& {Steele}}]{2009SmithB}
{Smith}, N., {Silverman}, J.~M., {Chornock}, R., {et~al.} 2009, \apj, 695, 1334

\bibitem[{{Stritzinger} {et~al.}(2012){Stritzinger}, {Taddia}, {Fransson},
  {Fox}, {Morrell}, {Phillips}, {Sollerman}, {Anderson}, {Boldt}, {Brown},
  {Campillay}, {Castellon}, {Contreras}, {Folatelli}, {Habergham}, {Hamuy},
  {Hjorth}, {James}, {Krzeminski}, {Mattila}, {Persson}, \&
  {Roth}}]{2012Stritzinger}
{Stritzinger}, M., {Taddia}, F., {Fransson}, C., {et~al.} 2012, \apj, 756, 173

\bibitem[{{Taddia} {et~al.}(2015){Taddia}, {Sollerman}, {Fremling},
  {Pastorello}, {Leloudas}, {Fransson}, {Nyholm}, {Stritzinger}, {Ergon},
  {Roy}, \& {Migotto}}]{2015A&A...580A.131T}
{Taddia}, F., {Sollerman}, J., {Fremling}, C., {et~al.} 2015, \aap, 580, A131

\bibitem[{{Taddia} {et~al.}(2013){Taddia}, {Stritzinger}, {Sollerman},
  {Phillips}, {Anderson}, {Boldt}, {Campillay}, {Castell{\'o}n}, {Contreras},
  {Folatelli}, {Hamuy}, {Heinrich-Josties}, {Krzeminski}, {Morrell}, {Burns},
  {Freedman}, {Madore}, {Persson}, \& {Suntzeff}}]{2013A&A...555A..10T}
{Taddia}, F., {Stritzinger}, M.~D., {Sollerman}, J., {et~al.} 2013, \aap, 555,
  A10

\bibitem[{Tielens(2005)}]{ISM}
Tielens, A. G. G.~M. 2005, The physics and chemistry of the interstellar medium
  (Cambridge University Press, Cambridge, UK)

\bibitem[{{Van Dyk}(2013)}]{VanDyk_2013AJ....145..118V}
{Van Dyk}, S.~D. 2013, \aj, 145, 118

\bibitem[{{van Marle} {et~al.}(2010){van Marle}, {Smith}, {Owocki}, \& {van
  Veelen}}]{2010vanMarle}
{van Marle}, A.~J., {Smith}, N., {Owocki}, S.~P., \& {van Veelen}, B. 2010,
  \mnras, 407, 2305

\bibitem[{{Wesson} {et~al.}(2015){Wesson}, {Barlow}, {Matsuura}, \&
  {Ercolano}}]{Wesson_2015MNRAS.446.2089W}
{Wesson}, R., {Barlow}, M.~J., {Matsuura}, M., \& {Ercolano}, B. 2015, \mnras,
  446, 2089

\end{thebibliography}

\end{document}